\documentclass{iaus}
\usepackage{graphics,graphicx}

  \checkfont{eurm10}
  \iffontfound
    \IfFileExists{upmath.sty}
      {\typeout{^^JFound AMS Euler Roman fonts on the system,
                   using the 'upmath' package.^^J}%
       \usepackage{upmath}}
      {\typeout{^^JFound AMS Euler Roman fonts on the system, but you
                   dont seem to have the}%
       \typeout{'upmath' package installed. iaus.cls can take advantage
                 of these fonts,^^Jif you use 'upmath' package.^^J}%
      }
  \else
  \fi

  \checkfont{msam10}
  \iffontfound
    \IfFileExists{amssymb.sty}
      {\typeout{^^JFound AMS Symbol fonts on the system, using the
                'amssymb' package.^^J}%
       \usepackage{amssymb}%

      }{}
  \fi

  \IfFileExists{amsbsy.sty}
    {\typeout{^^JFound the 'amsbsy' package on the system, using it.^^J}%
     \usepackage{amsbsy}}
    {}

\newsavebox{\astrutbox}
\sbox{\astrutbox}{\rule[-5pt]{0pt}{20pt}}

\newcommand{\niiH}{[N\,{\sc ii}]$\lambda6583/\mathrm{H}\alpha$}
\newcommand{\oiiiH}{[O\,{\sc iii}]$\lambda5007/\mathrm{H}\beta$}
\newcommand{\siiH}{[S\,{\sc ii}]$\lambda \lambda 6717,30/\mathrm{H}\alpha$}
\newcommand{\oiH}{[O\,{\sc i}]$\lambda6300/\mathrm{H}\alpha$}
\newcommand{\heiiH}{He\,{\sc ii}$\lambda4686$/H$\beta$}
\newcommand{\cii}{C\,{\sc ii}}
\newcommand{\ciii}{C\,{\sc iii}}
\newcommand{\civ}{C\,{\sc iv}}
\newcommand{\neiii}{Ne\,{\sc iii}}
\newcommand{\nev}{Ne\,{\sc v}}
\newcommand{\heii}{He\,{\sc ii}}

\title[Dusty Photoionization: The Solution to the NLR Problem]{Dusty, Radiation Pressure Dominated Photoionization: The Solution to the Narrow Line Region Problem}

\author[B. Groves, {\it et al.\/}]%
{Brent Groves$^1$,
Michael Dopita$^1$
\and Ralph Sutherland$^1$}

\affiliation{$^1$Research School of Astronomy \& Astrophysics, Australia National University, Canberra, Australia. email: bgroves@mso.anu.edu.au}

\pubyear{2004}
\volume{222}
\pagerange{1--8}
\date{?? and in revised form ??}
\jname{The Interplay among Black Holes, Stars and ISM \\in Galactic Nuclei}
\editors{Th. Storchi Bergmann, L.C. Ho \& H.R. Schmitt, eds.}
\begin{document}

\maketitle

\begin{abstract}
Seyfert narrow line regions (NLR) have emission line ratios which are remarkably
uniform, displaying only $\sim$0.5  dex variation between different galaxies.
Existing models have been unable to
explain  these observations without the introduction of ad hoc assumptions,
geometrical restrictions or new parameters. Here we introduce a new model:
dusty radiation pressure dominated photoionization, which provides
a natural self-regulating characteristic leading to an invariance of the
spectrum over a very wide range ($>100$) of ionization parameter. The dusty model is able to reproduce both the range and the
absolute value of the observational line ratios not only in the standard optical
diagnostic diagrams but also in UV diagnostic plots, providing an explanation to the problem in NLR observations.
\end{abstract}

\firstsection 
\section{Introduction}

The emission lines of active galaxies have often been used in
conjunction with models to constrain the physical and ionizati on
structure of the emitting regions.  In particular ratio diagrams or
line diagnostic diagrams prove to be an excellent visual aid in
interpreting the emission line data. For example, the line diagnostic
diagrams by \cite{VO87} are capable of distinguishing
three different groups of emission line galaxies: those excited by
starbursts and two excited by an active nucleus -
the Seyfert narrow line regions (NLRs) and the low ionization nuclear
emission-line regions (LINERs). These diagrams are additionally
interesting in that they show that the emission from Narrow Line Regions (NLR) is
remarkably uniform, with only $\sim$0.5 dex variation between Seyferts
and less within individual galaxies. This uniformity of the spectral
properties has since been confirmed in much larger samples
(eg.\ \cite{VeronCetty2000}).

The standard paradigm proposes that the NLR are excited by
photons originating at or near a compact nuclear source (see, eg.\ \cite{Ost89}) having a smooth featureless power-law, or broken power-law
EUV ionizing spectrum. Within this model, the clustering of the observed
line ratios within such a restricted domain of parameter space presents a
problem, as it requires an approximately
constant ionization parameter\footnote{The ionization parameter is a measure of the number of ionizing photons against the
hydrogen density ($U = S_{\star}/\mathrm{n}_{\mathrm{H}}c$).} of $U\sim10^{-2}$, whereas $U$ should be free to take on any value.  Modellers have been
therefore forced to make the arbitrary (and possibly unphysical) assumption
that the gas density in the ionized clouds must fall exactly as the inverse
square of the distance from the nucleus.

In order to account for this failings of the previous standard models we have
proposed a new paradigm for the photoionization of the NLR clouds,
that of dusty, radiation pressure dominated photoionization
(\cite{DG02}).

\section{A New Paradigm}

The inclusion of dust into photoionization models affects the final
emission spectrum in several ways. As well as simply absorbing EUV
radiation and competing with Hydrogen for the ionizing photons, dust
affects the temperature structure of the NLR clouds through the
process of photoelectric heating.  

In order to be physical, an isobaric photoionization model must
include the effects of radiation pressure. The force of radiation can
be imparted to both the gaseous medium and dust, and
results in a radiation pressure gradient. Since in photoionized nebulae dust grains are charged
and therefore locked to the plasma by coulomb forces,
the radiation pressure gradient on dust
results in a gas pressure gradient. Standard
photoionization models are isochoric and therefore cannot take this
effect into account.  To demonstrate these effects we have run
the standard isochoric model and the new dusty, radiation pressure
dominated model over a large set of input parameters.

With simple calculations it is easy to show that at an ionization
parameter of $\log U \sim -2$, dust begins to dominate the opacity of
the ionized cloud and hence the radiation pressure (\cite{Groves04a}). It is also around
this value of the ionization parameter that radiation pressure
developed at the ionization front of the NLR cloud becomes comparable
with the gas pressure.  Therefore at high ionization parameters, the
pressure in the ionized gas, and hence density, is determined by the external
ionization parameter, $U_0$ and the local ionization parameter becomes
independent of the external ionizing flux. The result of this is that
at high ionization parameter the emission line spectrum of the low-
and intermediate ionization species is effectively independent of the
external ionization parameter.

\section{Resulting Line Diagnostic Diagrams}

\begin{figure}
\includegraphics[width=65mm]{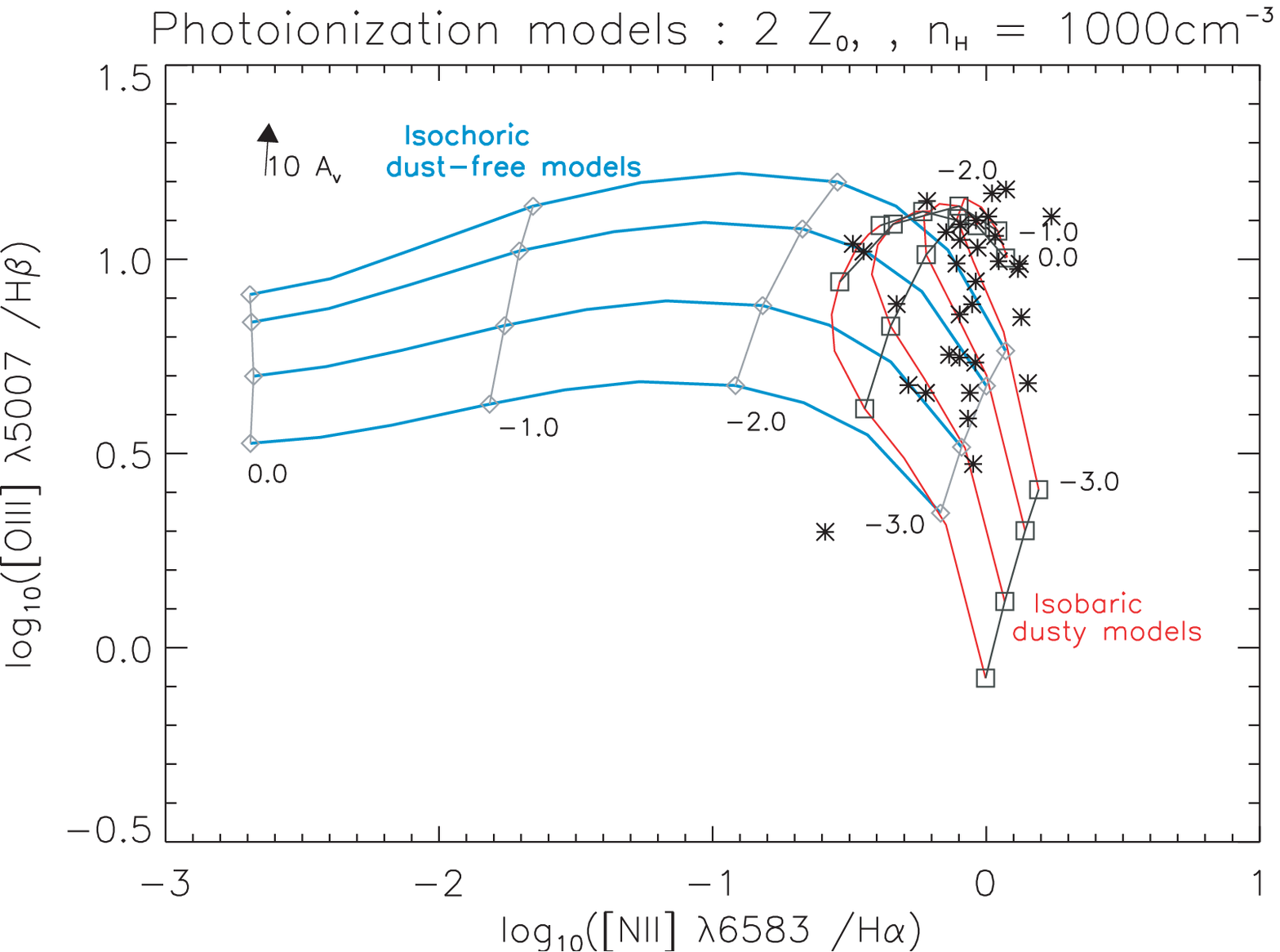}%
\hspace{5mm}
\includegraphics[width=65mm]{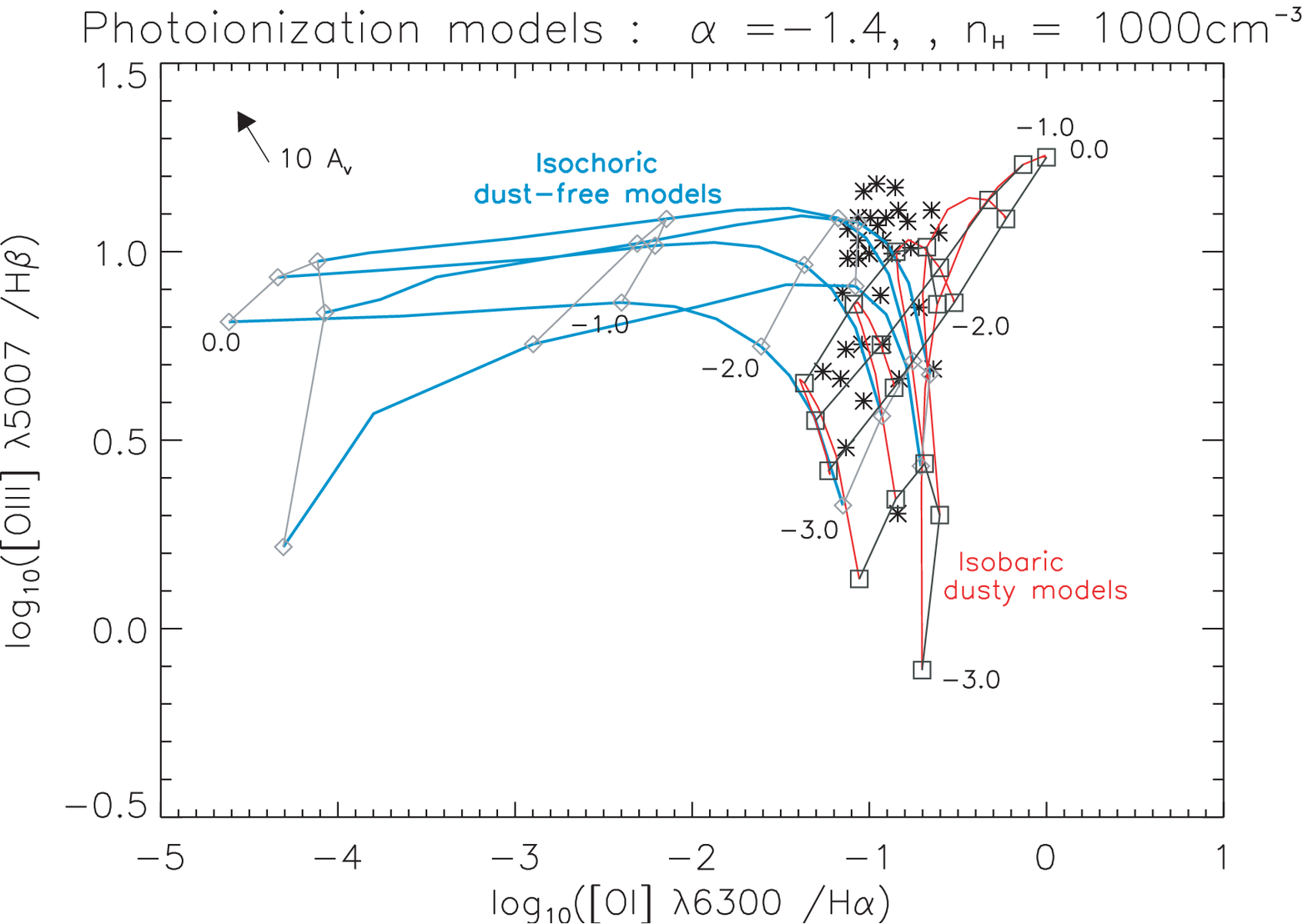}
\vspace{5mm}\\
\includegraphics[width=65mm]{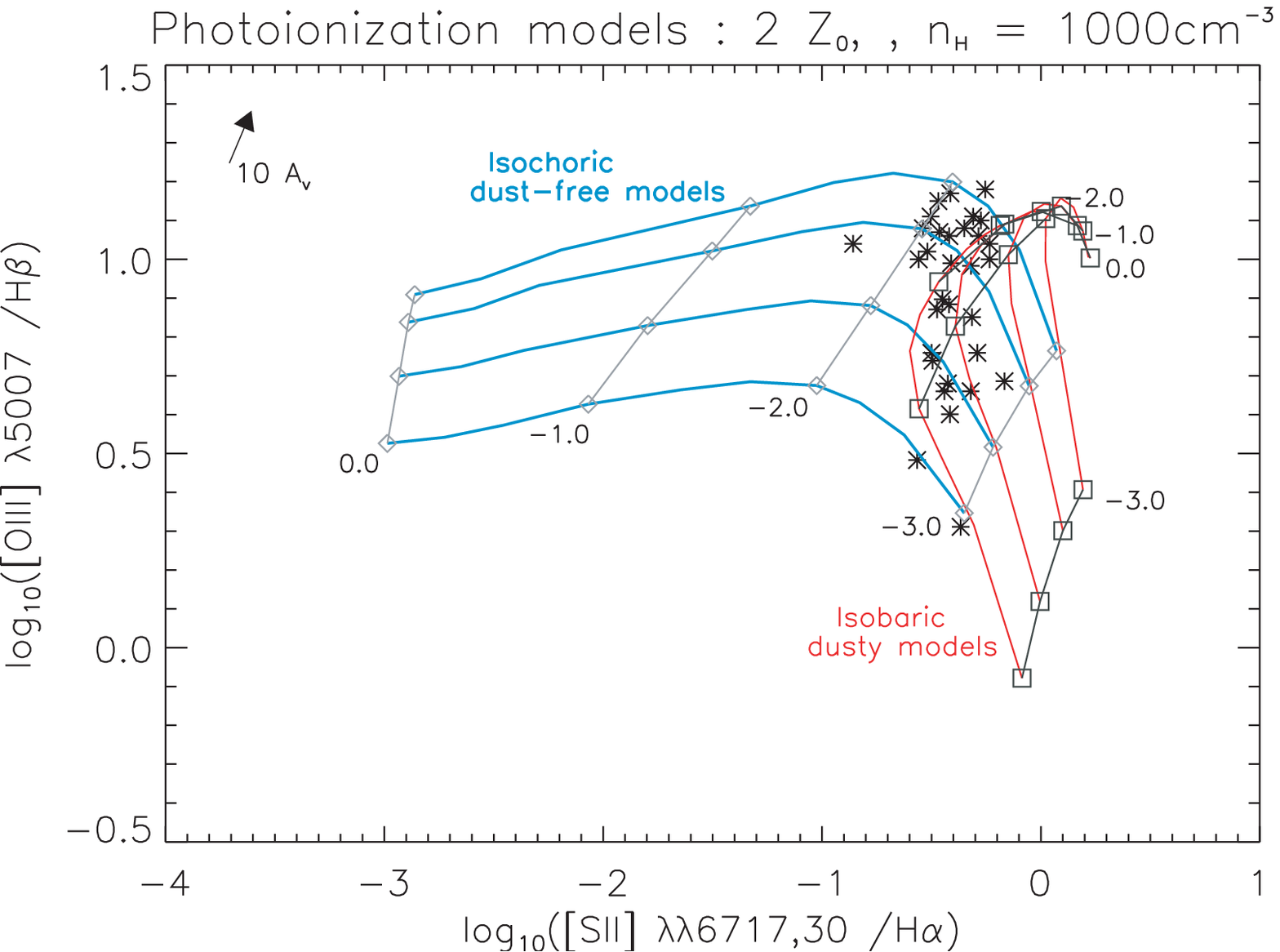}
\hspace{5mm}
\includegraphics[width=65mm]{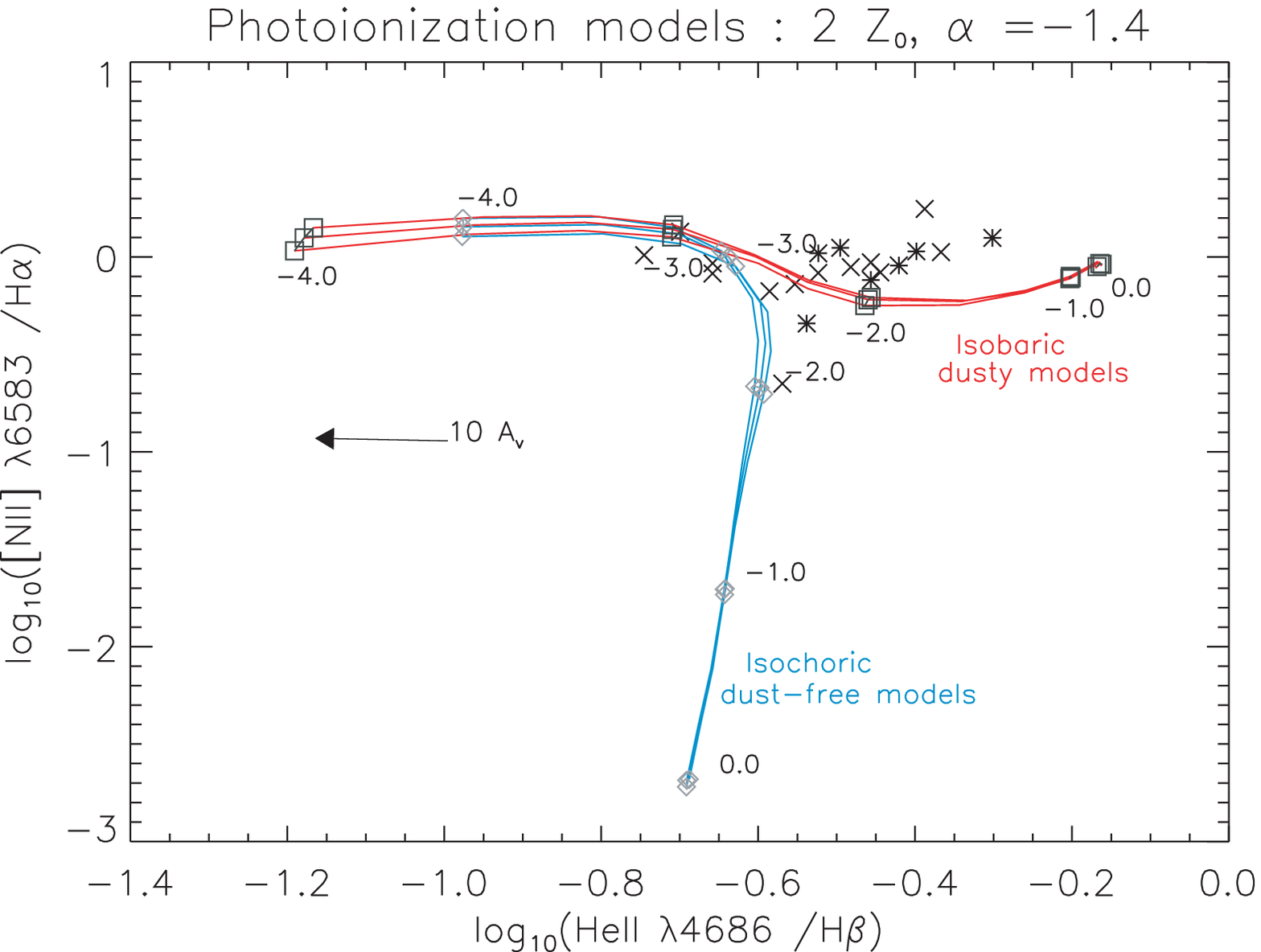}
\caption{\label{fig:vis}Four visible line diagnostic diagrams demonstrating the difference between a standard, dust-free photoionization model (light diamonds) and our dusty, radiation pressure dominated model (dark squares) in reproducing observations of seyfert galaxies (asterisks). On each figure the effects of 10 magnitudes of visual extinction due to dust are marked. Lines of constant ionization parameter are labelled with the value of $\log U_0$.}
\end{figure}

The success of the dusty models in reproducing a small range in standard emission line ratios over a large range in $U_0$ is clearly demonstrated in the standard line diagnostic diagrams shown in figure \ref{fig:vis} (\cite{Groves04b}).
Figure \ref{fig:vis}a, b \& c show the diagrams suggested by \cite{VO87} of \niiH, \oiH, and \siiH\ versus \oiiiH\ respectively. In each case the dust-free curves are able to reproduce the observations, yet are unable to properly constrain the data without assuming constant $U_0$. The dusty models however are able to not only reproduce the data but also "stagnates" at high $U_0$ in the region occupied by the observations.

Figure \ref{fig:vis}d shows the \heiiH\ vs. \oiiiH\ diagram which differs from the previous three as the He\,{\sc ii}/H$\beta$ ratio shows greater spread than the standard model. This demonstrates another success of the dusty models as they are able in this diagram to reproduce the spread in He\,{\sc ii}/H$\beta$ and the restricted range of [O\,{\sc iii}]/H$\beta$ in this  diagram. Here it is the hardening of the spectrum by dust and the determination of the density in the low-ionization region by radiation pressure which combine to create this spread in He\,{\sc ii}. 

\begin{figure}
\includegraphics[width=65mm]{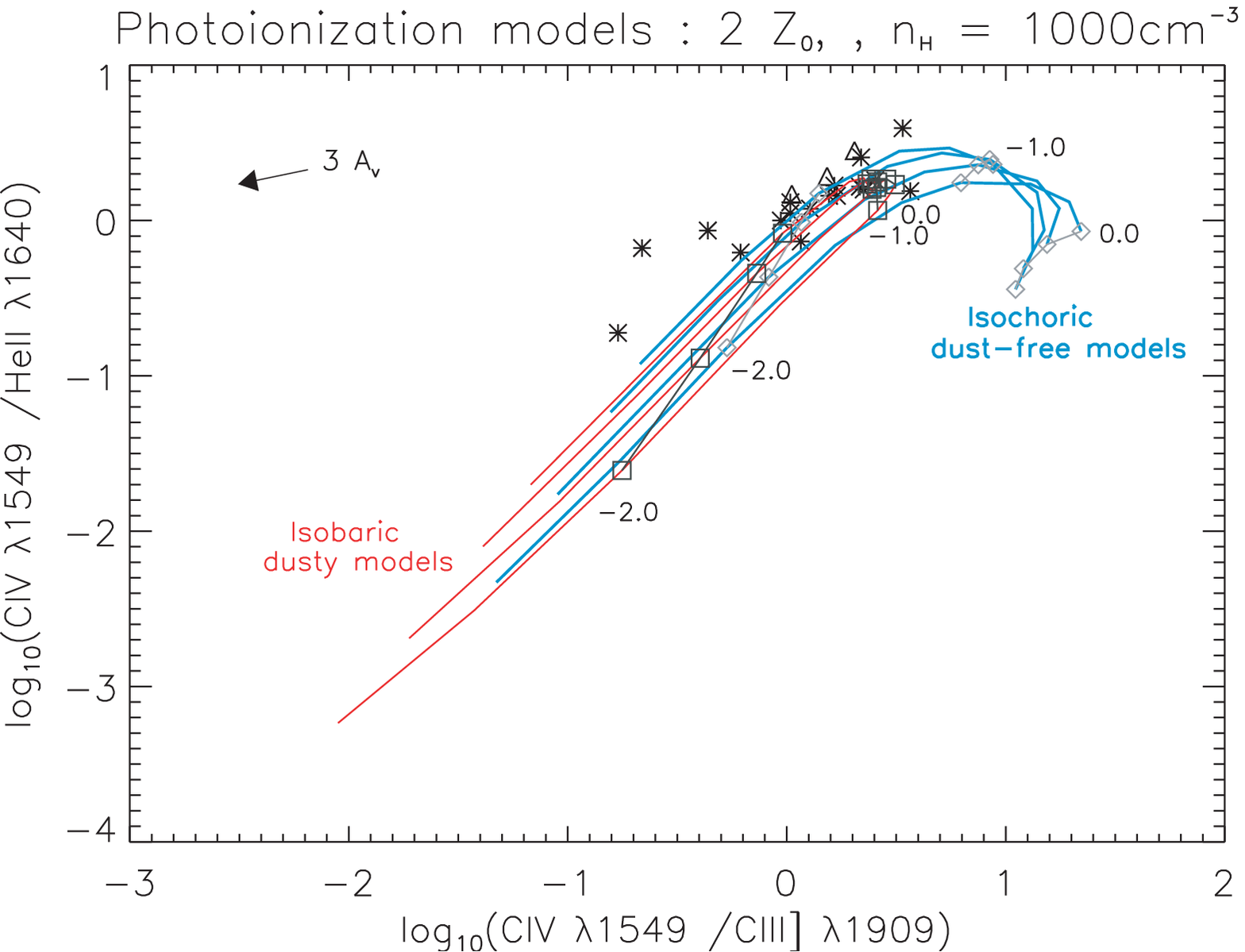}%
\hspace{5mm}
\includegraphics[width=65mm]{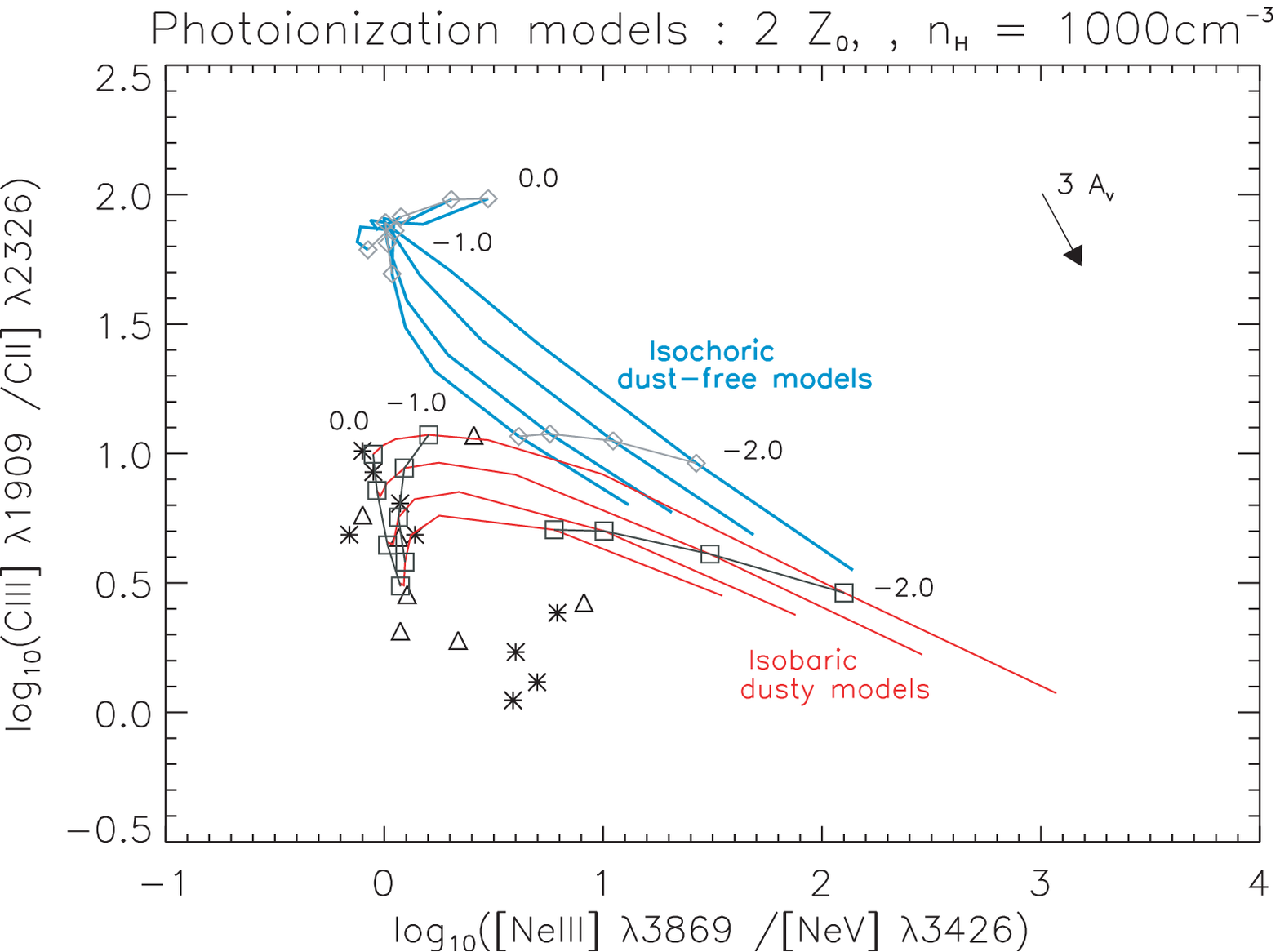}
\caption{\label{fig:UV} UV diagnostics demonstrating the success of the dusty radiation pressure dominated models in reproducing the observations compared to the standard dust-free photoionization models. Symbols are the same as in figure \ref{fig:vis}, except for the second figure where the observations are from high-$z$ radio galaxies.}
\end{figure}

The dusty model is not only able to reproduce the observed values and clustering of emission line ratios in the visible it is also able to do so at ultraviolet wavelengths as well. Figure \ref{fig:UV} demonstrates this with two UV line ratio diagnostic diagrams. The first, figure \ref{fig:UV}a, shows \civ\ $\lambda 1549$/\ciii]$\lambda 1909$ against \civ\ $\lambda 1549$/\heii
$\lambda 1640$, and reveals again the success of the dusty models in not only attain similar values to that observed in Seyfert galaxies, but in providing a mechanism for the observed clustering on the diagram.
The second figure, showing [\neiii]$\lambda3869$/[\nev]$\lambda 3426$ against \ciii]$\lambda 1909$/\cii]$\lambda 2326$, is an excellent diagnostic for shock excitation, and has been used as such for high-$z$ radio galaxies (\cite{Inskip02}). For those radio galaxies which are believed to be photoionized (high \ciii]/\cii]) the dusty models can reproduce the observed values, unlike the dust-free models, and the clustering at high $U_0$.

One detail stands out in all these diagnostics; the dusty,
radiation pressure dominated models provide undeniably the better fit
to the observations over the standard dust-free models. In all cases
the dusty models not only reproduce the data well, but also tend to
become degenerate in terms of the ionization parameter precisely in
the region occupied by the observations.

\section{Conclusion}

First introduced in \cite{DG02}, the dusty, radiation pressure
dominated photoionization model, through the stagnation of the ionization parameter at large values,
provides a simple explanation for the small variation of observed
Seyfert NLR ratios. This stagnation is due to the effects of dust opacity and radiation
pressure upon dust and is characteristic to these models. The
significant point is that the dusty model is able to do this over both
optical and UV ratios, without depending upon large variations in
other parameters such as density or metallicity. These results not only provide an explanation for what has not been a
fully understood observation for years but also provide ways in which
to understand further the processes involved in the NLR and extended
NLR of AGN.

\acknowledgements{
M.D.~acknowledges the support of the Australian National University
and of the Australian Research Council through his ARC Australian
Federation Fellowship and M.D.~and R.S.~acknowledges support through
the ARC Discovery project DP0208445.}

\end{document}